\documentclass[12pt]{article}

\usepackage[english]{babel}

\usepackage{latexsym}
\usepackage{amssymb}
\usepackage{epsfig}

\begin{document}

\title{Electronic states and transport properties in the Kronig-Penney
model with correlated compositional and structural disorder}

\author{ J. C. Hern\'{a}ndez-Herrej\'{o}n${}^{1}$, F.~M.~Izrailev${}^{2}$,
and L.~Tessieri${}^{1}$ \\
{\it ${}^{1}$ Instituto de F\'{\i}sica y Matem\'{a}ticas} \\
{\it Universidad Michoacana de San Nicol\'{a}s de Hidalgo} \\
{\it Morelia, Mich., 58060, Mexico} \\
{\it ${}^{2}$ Instituto de F\'{\i}sica, Universidad Aut\'{o}noma de Puebla,} \\
{\it Puebla, Pue., 72570, Mexico}}

\date{15th March 2010}

\maketitle

\begin{abstract}
We study the structure of the electronic states and the transport properties
of a Kronig-Penney model with weak compositional and structural disorder.
Using a perturbative approach we obtain an analytical expression
for the localisation length which is valid for disorder with arbitrary
correlations.
We show how to generate disorder with self- and cross-correlations and we
analyse both the known delocalisation effects of the long-range
self-correlations and new effects produced by cross-correlations.
We finally discuss how both kinds of correlations alter the
transport properties in Kronig-Penney models of finite size.
\end{abstract}

Pacs: 73.20.Jc, 73.20.Fz, 71.23.An

\section{Introduction}

The recent surge of interest in one-dimensional disordered models
can be attributed in large part to the realisation that long-range
correlations of the random potentials can significantly alter the
structure of the electronic states, enhancing or suppressing
their localisation in continuous intervals of
energy~\cite{Mou98,Izr99,Izr01,Kuh00,Kuh08}.
This discovery has paved the way for the construction of one-dimensional
electronic, optical or electromagnetic devices with anomalous transport
properties. Specifically, it is now possible to design filters with almost
perfect transmission  or reflection in pre-defined energy windows.
This possibility, first foreseen at the theoretical level~\cite{Izr99,Izr01},
was later confirmed experimentally~\cite{Kuh00,Kuh08}.

The periodic Kronig-Penney model was originally introduced in the early
1930s to analyse the electronic states and the energy bands in crystalline
structures~\cite{Kro31}. In the 1980s the model was used to describe
semiconductor superlattices~(see~\cite{Dav98} and references therein).
More recently, aperiodic variants of the Kronig-Penney model were used
to analyse the transmission properties of a waveguide with long-range
correlated compositional~\cite{Kuh00,Kuh08} or structural~\cite{Lun09}
disorder.
The wide applicability of aperiodic Kronig-Penney models makes desirable
to analyse in full detail the structure of their electronic states. Two
kinds of disorder are typically considered in the aperiodic Kronig-Penney
model: structural and compositional disorder.
In the first case the positions of potential barriers are randomised;
thus, they do not coincide with the sites of the underlying lattice.
In the second case the disorder is introduced by assigning random heights
or widths to the potential barriers.

In this paper we study a Kronig-Penney model with delta-shaped potential
barriers with weak correlated disorder of both the compositional and
structural type.
This model was first analysed in~\cite{Her08}, where we presented an
analytical expression for the inverse localisation length valid for
any kind of self- and cross-correlations of the two disorders.
Here we give a detailed derivation of this result, together with an
extended discussion of various physical implications.
In particular, we describe how to construct two sequences of random
variables, representing the positions and strengths of the barriers,
with arbitrary self- and cross-correlations.
We show that specific long-range self-correlations can create a
delocalisation-localisation transition (within the limit of the
second-order perturbative approach) and we analyse new effects
produced by the cross-correlations of the two kinds of disorder.
We then discuss how the both kinds of correlations affect
the electronic transport properties in finite samples.

This paper is organised as follows. In Sec.~\ref{ses} we define
the model under study and derive an analytical expression for the
electronic localisation length using the Hamiltonian map approach.
In Sec.~\ref{dme} we show how to construct sequences of correlated
random variables for the positions and strengths of the barriers.
We also demostrate how the localisation/delocalisation effects produced
by long-range self-correlations of the disorder are modified by the
cross-correlations.
In Sec.~\ref{trans} we analyse how these effects alter the transport
properties of finite-size models. The conclusions are drawn in
Sec.~\ref{conclu}.

\section{Structure of the electronic states}
\label{ses}

\subsection{Definition of the model}

Here we define the Kronig-Penney model with weak compositional and structural
disorder. The model describes the motion of an electron in a succession of
delta-shaped barriers; disorder is introduced by assuming that both the
strength and the position of each barrier are random variables fluctuating
around their average values. The corresponding Schr\"{o}dinger equation is
\begin{equation}
-\psi^{\prime \prime}(x) + \left[ \sum_{n=-\infty}^{\infty} U_{n}
\delta(x - d_{n}) \right] \psi(x) = q^{2} \psi(x)
\label{kpmodel}
\end{equation}
with the electron energy given by the square of the propagation wavevector
within the wells
\begin{equation}
E = q^{2}.
\label{eq2}
\end{equation}
In Eqs.~(\ref{kpmodel}) and~(\ref{eq2}) we have used units such that
$\hbar^{2}/2m = 1$; we will stick to this choice throughout this paper.

In the model~(\ref{kpmodel}) the delta-barriers are located at the points
$d_{n} = na + a_{n}$, where the random shifts $a_{n}$ with respect to the
lattice sites $na$ represent the structural (or {\em positional}) disorder.
The compositional (or {\em amplitude}) disorder, on the other hand, enters
via $U_{n} = U + u_{n}$, with the random variables $u_{n}$ corresponding to
fluctuations of the barrier strengths around their mean value $U$.

We define the statistical properties of the model in terms of the strength
fluctuations $u_{n}$ and of the relative barrier displacements
$\Delta_{n} = a_{n+1} - a_{n}$ (which turn out to be more relevant than
the absolute displacements $a_{n}$ themselves).
We restrict our attention to the weak-disorder case, defined by the
conditions
\begin{equation}
\langle u_{n}^{2} \rangle \ll U^{2}
\label{ucond}
\end{equation}
and
\begin{eqnarray}
\langle \Delta_{n}^{2} \rangle q^{2} \ll 1 & \mbox{ and } &
\langle \Delta_{n}^{2} \rangle U \ll 1 .
\label{dcond}
\end{eqnarray}
The condition~(\ref{ucond}) implies that the compositional disorder is weak
in the sense that the fluctuations of the barrier strength are small with
respect to their mean value.
The weakness of structural disorder is defined by the condition~(\ref{dcond}),
which requires that the relative displacements of the barriers should be
small on the length scales set by $1/q$ and $1/\sqrt{U}$.

In the weak-disorder case the statistical properties of the model can be
analysed in terms of the averages and of the binary correlators of the
random variables $u_{n}$ and $\Delta_{n}$. Both variables are assumed to
have zero mean.
In what follows we consider the normalised binary correlators
\begin{equation}
\begin{array}{ccl}
\chi_{1} (k) & = & \displaystyle
\frac{\langle u_{n} u_{n+k} \rangle}{\langle u_{n}^{2} \rangle} \\
\chi_{2} (k) & = & \displaystyle
\frac{\langle \Delta_{n} \Delta_{n+k} \rangle}
{\langle \Delta_{n}^{2} \rangle} \\
\chi_{3} (k) & = & \displaystyle
\frac{\langle u_{n} \Delta_{n+k} \rangle} {\langle u_{n} \Delta_{n} \rangle} \\
\end{array}
\label{bincor}
\end{equation}
as given functions. We do not attribute any specific form to the
correlators~(\ref{bincor}); we only suppose that they are even
functions of the index difference as a consequence of the assumed spatial
isotropy and homogeneity in the mean of the model.

\subsection{The Hamiltonian map}

The electronic states of the Kronig-Penney model~(\ref{kpmodel}) can
be analysed by means of the Hamiltonian map approach.
This approach is based on the mathematical identity of the
Schr\"{o}dinger equation~(\ref{kpmodel}) with the dynamical equation of
a classical oscillator with noisy frequency~\cite{Izr95,Izr98}.
The stochastic oscillator corresponding to the Kronig-Penney model is
defined by the Hamiltonian
\begin{equation}
H = \frac{p^{2}}{2} + \frac{1}{2} \left\{ q^{2} -
\left[ \sum_{n=-\infty}^{\infty} U_{n} \delta(t - na + a_{n}) \right] \right\}
x^{2} .
\label{kickosc}
\end{equation}
with the electron wavevector $q$ playing the role of the unperturbed
frequency.
The mathematical analogy between the Schr\"{o}dinger
equation~(\ref{kpmodel}) and the dynamical equation of the parametric
oscillator~(\ref{kickosc}) allows one to analyse the behaviour of
the electronic states of the Kronig-Penney model in terms of the
trajectories of the kicked oscillator.
This dynamical approach implies that the Schr\"{o}dinger equation is solved
as an initial-value problem; hence the Hamiltonian map method is
equivalent to the transfer matrix technique.

Given the Hamiltonian~(\ref{kickosc}), one can integrate the corresponding
dynamical equations over the time interval $[d_{n}^{-}, d_{n+1}^{-}]$ between
two kicks. In this way one obtains the Hamiltonian map
\begin{equation}
\left( \begin{array}{c}
x_{n+1} \\
p_{n+1} \\
\end{array} \right) =
{\bf T}_{n}
\left( \begin{array}{c}
x_{n} \\
p_{n} \\
\end{array} \right)
\label{hammap1}
\end{equation}
with the transfer matrix
\begin{equation}
{\bf T}_{n} = \left(
\begin{array}{cc}
\displaystyle
\cos \left[q \left( a + \Delta_{n} \right)\right] +
\left( U + u_{n} \right) \frac{1}{q}
\sin \left[ q \left( a + \Delta_{n} \right)\right] &
\displaystyle
\frac{1}{q} \sin \left[q \left( a + \Delta_{n} \right) \right]\\
\displaystyle
-q \sin \left[q \left( a + \Delta_{n} \right) \right]
+\left( U + u_{n} \right)
\cos \left[ q \left( a + \Delta_{n} \right) \right]&
\displaystyle
\cos \left[ q \left( a + \Delta_{n} \right) \right]\\
\end{array}
\right) .
\label{transfer}
\end{equation}
To analyse the trajectories of the Hamiltonian map~(\ref{hammap1}), we
follow the approach proposed in~\cite{Izr01}.
Specifically we perform a canonical transformation
$(x_{n}, p_{n}) \rightarrow (X_{n}, P_{n})$ such that the
unperturbed motion reduces to a simple rotation in the new variables.
The appropriate canonical transformation has the form
\begin{equation}
\left( \begin{array}{c}
x_{n} \\
p_{n} \\
\end{array} \right) =
{\bf M}
\left( \begin{array}{c}
X_{n} \\
P_{n} \\
\end{array} \right)
\label{canonic}
\end{equation}
with
\begin{displaymath}
{\bf M} = \left( \begin{array}{cc}
\displaystyle
\alpha \cos \frac{q a}{2} &
\displaystyle
\frac{1}{q \alpha} \sin
\frac{q a}{2} \\
\displaystyle
- q \alpha \sin \frac{q a}{2} &
\displaystyle
\frac{1}{\alpha} \cos \frac{q a}{2} \\
\end{array} \right)
\end{displaymath}
where the parameter $\alpha$ is defined by the relation
\begin{equation}
\alpha^{4} = \frac{1}{q^{2}}
\frac{\displaystyle \sin \left(q a \right)
- \frac{U}{2q} \left[ \cos \left( q a \right) - 1 \right]}
{\displaystyle
\sin \left(q a \right) - \frac{U}{2q}
\left[ \cos \left( q a \right) + 1 \right]} .
\label{alpha}
\end{equation}
Note that the canonical transformation~(\ref{canonic}) rescales time so
that the new coordinate $X_{n}$ and the corresponding conjugate momentum
$P_{n}$ have the same dimensions.

Since we are interested only in the weak-noise case, we can expand the new
transfer matrix $\tilde{\bf T}_{n} = {\bf M}^{-1} {\bf T}_{n} {\bf M}$
and keep only terms up to the second order.
The Hamiltonian map then takes the form
\begin{equation}
\left( \begin{array}{c}
X_{n+1} \\
P_{n+1} \\
\end{array} \right) =
\left[ \tilde{\bf T}_{n}^{(0)} + \tilde{\bf T}_{n}^{(1)} +
\tilde{\bf T}_{n}^{(2)} +o(2) \right]
\left( \begin{array}{c}
X_{n} \\
P_{n} \\
\end{array} \right)
\label{hammap3}
\end{equation}
where, as desired, the unperturbed matrix represents a rotation
\begin{displaymath}
\tilde{\bf T}_{n}^{(0)} =
{\bf M}^{-1} {\bf T}_{n}^{(0)} {\bf M} =
\left( \begin{array}{cc}
\cos \left( k a \right) & \sin \left( k a \right) \\
-\sin \left( k a \right) & \cos \left( k a \right) \\
\end{array} \right),
\end{displaymath}
with the rotation angle $k a$ being defined by the equation
\begin{equation}
\cos \left( k a \right) = \cos \left( q a \right) +
\frac{U}{2q} \sin \left( q a \right) .
\label{rotangle}
\end{equation}
In terms of the Kronig-Penney model, $k$ is the Bloch wavevector,
while equation~(\ref{rotangle}) defines the band structure of the
model.
The explicit form of the first- and the second-order terms of the
map~(\ref{hammap3}) is
\begin{displaymath}
\tilde{\bf T}_{n}^{(1)} =
\left( \begin{array}{cc}
\displaystyle
\frac{\sin \left( qa \right)}{2q} u_{n} -
\frac{1}{\alpha^{2}} \sin \left( ka \right) \Delta_{n} &
\displaystyle
\frac{1}{\alpha^{2}} \left[ \frac{1 - \cos \left( qa \right)}{2 q^{2}} u_{n}
- \cos \left( ka \right) \Delta_{n} \right] \\
\displaystyle
\alpha^{2} q^{2} \left[ \frac{1 + \cos \left( qa \right)}{2 q^{2}} u_{n} -
\cos \left( ka \right) \Delta_{n} \right] &
\displaystyle
\frac{\sin \left( qa \right)}{2q} u_{n} -
\alpha^{2} q^{2} \sin \left( ka \right) \Delta_{n} \\
\end{array}
\right)
\end{displaymath}
and
\begin{displaymath}
\tilde{\bf T}_{n}^{(2)} = \left( \begin{array}{cc}
\displaystyle
\frac{\cos \left( qa \right) + 1}{2} u_{n} \Delta_{n} -
\frac{1}{2} q^{2} \cos \left( ka \right) \Delta_{n}^{2} &
\displaystyle
\frac{\sin \left( qa \right)}{2 q \alpha^{2}} u_{n} \Delta_{n} -
\frac{1}{2} q^{2} \sin \left( ka \right) \Delta_{n}^{2} \\
\displaystyle
-\frac{q \alpha^{2}}{2} \sin \left( qa \right) u_{n} \Delta_{n} +
\frac{1}{2} q^{2} \sin \left( ka \right) \Delta_{n}^{2} &
\displaystyle
\frac{\cos \left( qa \right) - 1}{2} u_{n} \Delta_{n} -
\frac{1}{2} q^{2} \cos \left( ka \right) \Delta_{n}^{2} \\
\end{array}
\right) .
\end{displaymath}

A remark is in order here. The canonical transformation~(\ref{canonic}) is
well-defined for every value of the rotation angle $k a$ {\em
except} for the critical values $k a = 0$ and $k a
= \pm \pi$. In terms of the Kronig-Penney model, this means that
our approach fails at the centre and at the edges of the first Brillouin
zone. This can be seen by considering that Eq.~(\ref{rotangle}) implies
that
\begin{displaymath}
\sin \left( k a \right) = \pm \sqrt{ \left\{ \sin \left( q a
\right) - \frac{U}{2q} \left[ \cos \left( q a \right) + 1
\right] \right\} \left\{ \sin \left( q a \right) - \frac{U}{2q}
\left[ \cos \left( q a \right) - 1 \right] \right\}}.
\end{displaymath}
This identity shows that, whenever the sine of the rotation angle is equal
to zero, the $\alpha$ parameter~(\ref{alpha}) either vanishes or diverges.
In both cases the canonical transformation~(\ref{canonic}) is not properly
defined, because some elements of the matrix ${\bf M}$ diverge. However,
we would like to stress that the canonical transformation is perfectly
well-defined as long as the rotation angle is arbitrarily close (but not
exactly identical) to the singular values $k a = 0$ and $k a
= \pm \pi$. Hence our approach works well in every neighbourhood of these
critical points.

To analyse the evolution of the dynamical system~(\ref{hammap3}), it is
convenient to switch from Cartesian to action-angle coordinates, defined
via the equations
\begin{displaymath}
\begin{array}{ccl}
X_{n} & = & \sqrt{2 J_{n}} \sin \theta_{n} \\
P_{n} & = & \sqrt{2 J_{n}} \cos \theta_{n} \\
\end{array} .
\end{displaymath}
In terms of the new variables, the Hamiltonian map~(\ref{hammap3}) takes
the form
\begin{equation}
\begin{array}{ccl}
J_{n+1} & = & D_{n}^{2} J_{n} \\
\theta_{n+1} & = & \displaystyle
\theta_{n} + ka - \frac{1}{2}
\left[1 - \cos \left( 2 \theta_{n} + ka \right) \right] \tilde{u}_{n} 
+ \frac{1}{2} \left[ \zeta - \cos \left( 2 \theta_{n} + 2 ka \right) \right]
\tilde{\Delta}_{n} \\
& + & \displaystyle
\frac{1}{8}
\left[ 2 \sin \left( 2 \theta_{n} + ka \right) - \sin \left( 4 \theta_{n}
+ 2 ka \right) \right] \tilde{u}_{n}^{2} \\
& + & \displaystyle
\frac{1}{8} \left[ 2 \zeta \sin \left( 2 \theta_{n} + 2 ka \right) -
\sin \left( 4 \theta_{n} + 4 ka \right) \right] \tilde{\Delta}_{n}^{2} \\
& + & \displaystyle
\frac{1}{4} \left[ \sin \left( ka \right) -
2 \sin \left( 2 \theta_{n} + 2 ka \right) + \sin \left( 4 \theta_{n} +
3 ka \right) \right] \tilde{u}_{n} \tilde{\Delta}_{n} ,
\end{array}
\label{hammap5}
\end{equation}
with the ratio of the action variables being equal to
\begin{equation}
\begin{array}{ccl}
D_{n}^{2} & = & \displaystyle
1 + \sin \left( 2 \theta_{n} + ka \right) \tilde{u}_{n} -
\sin \left( 2 \theta_{n} + 2 ka \right) \tilde{\Delta}_{n} \\
& + & \displaystyle
\frac{1}{2} \left[ 1 - \cos \left( 2 \theta_{n} + k a \right) \right]
\tilde{u}_{n}^{2} +
\frac{1}{2} \left[ 1 - \zeta \cos \left( 2 \theta_{n} + 2 k a \right)
\right] \tilde{\Delta}_{n}^{2} \\
& + & \displaystyle
\left[ \cos \left( 2 \theta_{n} + 2 k a \right) -\cos \left( k a \right)
\right] \tilde{u}_{n} \tilde{\Delta}_{n} .
\end{array}
\label{dnsquared}
\end{equation}
In Eqs.~(\ref{hammap5}) and~(\ref{dnsquared}) we made use of the
rescaled random variables
\begin{eqnarray}
\tilde{u}_{n} =
\frac{\sin \left( qa \right)}{q \sin \left( ka \right)} u_{n} &
\mbox{ and } &
\tilde{\Delta}_{n} = \frac{U}{\sin \left( ka \right)} \Delta_{n} 
\label{rescaled}
\end{eqnarray}
and we introduced the short-hand notation
\begin{equation}
\zeta = \frac{q \sin \left( ka \right)}{U} \left[ q \alpha^{2} +
\frac{1}{q \alpha^{2}} \right] .
\label{zeta}
\end{equation}
We remark that the angle variable evolves independently of the action
variable.

\subsection{The localisation length}

This subsection is devoted to the task of evaluating the localisation
length in the Kronig-Penney model~(\ref{kpmodel}).
The inverse localisation length is defined as
\begin{displaymath}
l^{-1}_{\rm loc} = \lim_{N \rightarrow \infty} \frac{1}{Na} \sum_{n=1}^{N}
\log \left|\frac{\psi_{n+1}}{\psi_{n}}\right|
\end{displaymath}
which, in dynamical terms, is equivalent to the Lyapunov exponent of the
Hamiltonian map~(\ref{hammap3}), i.e.,
\begin{equation}
\lambda = \lim_{N \rightarrow \infty} \frac{1}{2Na} \sum_{n=1}^{N}
\log \left( \frac{J_{n+1}}{J_{n}} \right)
= \frac{1}{2a} \langle \log D_{n}^{2} \rangle .
\label{lyap1}
\end{equation}
To compute the Lyapunov exponent~(\ref{lyap1}) we expand the logarithm;
within the second-order approximation one obtains
\begin{equation}
\begin{array}{ccl}
\lambda & = & \displaystyle
\frac{1}{2a} \biggl\langle
\sin \left( 2 \theta_{n} + ka \right) \tilde{u}_{n} -
\sin \left( 2 \theta_{n} + 2 ka \right) \tilde{\Delta}_{n} \\
& + & \displaystyle
\frac{1}{4} \left[ 1 - 2 \cos \left( 2 \theta_{n} + ka \right) +
\cos \left( 4 \theta_{n} + 2 ka \right) \right] \tilde{u}_{n}^{2} \\
& + & \displaystyle
\frac{1}{4} \left[ 1 - 2 \zeta \cos \left( 2 \theta_{n} + 2 ka \right)
+ \cos \left( 4 \theta_{n} +  4ka \right) \right] \tilde{\Delta}_{n}^{2} \\
& - & \displaystyle
\frac{1}{2} \left[ \cos \left( ka \right) - 2 \cos \left( 2 \theta_{n} +
2 ka \right) + \cos \left( 4 \theta_{n} + 3 ka \right) \right]
\tilde{u}_{n} \tilde{\Delta}_{n}  \biggr\rangle \\
\end{array}
\label{logd}
\end{equation}
The terms in the angular brackets of the right-hand side (rhs) of
Eq.~(\ref{logd}) must be averaged over the angle variable with an
appropriate distribution $\rho(\theta)$.
Typically, one uses a flat distribution $\rho (\theta) = 1/(2 \pi)$.
In fact, it is easy to see that in the absence of disorder the
Hamiltonian map~(\ref{hammap5}) reduces to the form
\begin{displaymath}
\begin{array}{ccl}
J_{n+1} & = & J_{n} \\
\theta_{n + 1} & = & \theta_{n} + k a .\\
\end{array}
\end{displaymath}
Hence the angle variable has a fast dynamics compared to the action
variable and quickly assumes a uniform distribution of values in the
interval $[0:2 \pi]$.

This reasoning fails, however, when the rotation angle is either zero
or a rational multiple of $\pi$. In the latter case the noiseless
angular map has periodic orbits which manifest themselves in the form of
a slight modulation of the invariant measure when a weak noise is switched
on. In particular, when one has
\begin{displaymath}
k a = \frac{\pi}{n}
\end{displaymath}
the uniform distribution of the angular variable is modified by a
perturbative term proportional to $\cos (2n \theta)$ or $\sin (2 n \theta)$.
This modulation of the invariant measure alters the value of the
inverse localisation length~(\ref{logd}) only for $n = \pm 1$, and
$n = \pm 2$, because only second- and fourth-order harmonics are present
in the rhs of Eq.~(\ref{logd}).
The values $ka =0$ and $ka = \pm \pi$ correspond to the centre and the edges
of the first Brillouin zone, i.e., to the edges of the energy bands. The
values $ka = \pm \pi/2$ of the Bloch vector, on the other hand, correspond
to values of the energy close to the band centre.
For these special values of the rotation angle the assumption of a flat
invariant distribution must be abandoned and the specific form of
$\rho(\theta)$ has to be determined before one can compute the average in
Eq.~(\ref{logd}). The non-uniform distribution of the angular variable
produces anomalies in the localisation length; we will analyse them
elsewhere.

Except for the special cases mentioned above, one can compute the averages
in Eq.~(\ref{logd}) with a uniform angular distribution.
The corresponding result is
\begin{equation}
\begin{array}{ccl}
\lambda & = & \displaystyle
\frac{1}{8a} \left[ \langle \tilde{u}_{n}^{2} \rangle +
\langle \tilde{\Delta}_{n}^{2} \rangle -
2 \langle \tilde{u}_{n} \tilde{\Delta}_{n} \rangle \cos \left( ka \right)
\right] \\
& + & \displaystyle
\frac{1}{2a} \left[ \langle \tilde{u}_{n} \sin \left( 2 \theta_{n}
+ ka \right) \rangle - \langle \tilde{\Delta}_{n} \sin \left( 2 \theta_{n}
+ 2 ka \right) \rangle \right] .\\
\end{array}
\label{lyap2}
\end{equation}
To proceed further, we need to compute the noise-angle correlators which
appear in formula~(\ref{lyap2}). One can generalise the method used
in~\cite{Izr99} and introduce the correlator between the kick strength
and the angle
\begin{displaymath}
r_{l} = \langle \tilde{u}_{n} \exp \left( i 2 \theta_{n-l} \right) \rangle
\end{displaymath}
as well as the correlator between the kick timing and the angle
\begin{displaymath}
s_{l} = \langle \tilde{\Delta}_{n}
\exp \left( i 2 \theta_{n-l} \right) \rangle .
\end{displaymath}
Both correlators satisfy recursive relations, which can be derived
as follows.
Keeping only second-order terms in the disorder strength, one can
write $r_{l-1}$ in the form
\begin{equation}
\begin{array}{ccl}
r_{l-1} & = & \displaystyle
\Big\langle \tilde{u}_{n} \exp \left( i 2 \theta_{n-l+1} \right) \Big\rangle
= \Big\langle \tilde{u}_{n} \exp \left( i 2 \theta_{n-l} \right)
\exp \left( i 2 k a \right) \\
& \times & \displaystyle
\left\{ 1 - i \left[ 1 - \cos \left( 2 \theta_{n-l} + ka \right) \right]
\tilde{u}_{n-l}
+ i \left[ \zeta - \cos \left( 2 \theta_{n-l} + 2 ka \right) \right]
\tilde{\Delta}_{n-l} \right\} \Big\rangle . \\
\end{array}
\label{rcorr}
\end{equation}
For weak disorder the triple and quadruple noise-angle correlators can
be factorised and averages over the angle variable can be carried out with
a uniform distribution. The result is
\begin{displaymath}
r_{l-1} = r_{l} \exp \left( i 2 k a \right) +
\frac{i}{2} \exp \left( i ka  \right) \langle \tilde{u}_{n} \tilde{u}_{n-l}
\rangle - \frac{i}{2} \langle \tilde{u}_{n} \tilde{\Delta}_{n-l} \rangle .
\end{displaymath}
Multiplying both members of this identity by $\exp \left[ i 2 k a
\left( l - 1 \right) \right]$ and summing over $l$ from zero to infinity,
one obtains
\begin{displaymath}
r_{0} = \frac{i}{2} \exp \left(-ika \right) \sum_{l=1}^{\infty}
\langle \tilde{u}_{n} \tilde{u}_{n-l} \rangle \exp \left( i2kal \right) -
\frac{i}{2} \exp \left( -i2ka \right) \sum_{l=1}^{\infty}
\langle \tilde{u}_{n} \tilde{\Delta}_{n-l} \rangle \exp \left( i2kal \right).
\end{displaymath}
From this expression one can easily compute the correlator between the
angle and the kick strength in Eq.~(\ref{lyap2}); in fact, one has
\begin{equation}
\begin{array}{ll}
& \displaystyle
\langle \tilde{u}_{n} \sin \left( 2 \theta_{n} + k a \right) \rangle =
{\rm Im} \left[ r_{0} \exp \left( i k a \right) \right] \\
= & \displaystyle
\frac{1}{2} \sum_{l=1}^{\infty} \langle \tilde{u}_{n} \tilde{u}_{n-l} \rangle
\cos \left( 2 k a l \right) - \frac{1}{2}
\sum_{l=1}^{\infty} \langle \tilde{u}_{n} \tilde{\Delta}_{n-l} \rangle
\cos \left[ k a \left( 2l -1 \right) \right] .
\end{array}
\label{uthcorr}
\end{equation}
Following the same approach, one obtains that the correlator between the
angle and the timing of the kick is
\begin{equation}
\begin{array}{ll}
& \displaystyle
\langle \tilde{\Delta}_{n} \sin \left( 2 \theta_{n} + 2 k a \right) \rangle
= {\rm Im} \left[ s_{0} \exp \left( i 2 k a \right) \right] \\
= & \displaystyle
\frac{1}{2} \sum_{l=1}^{\infty} \langle \tilde{\Delta}_{n} \tilde{u}_{n-l}
\rangle \cos \left[ k a \left( 2 l + 1 \right) \right] -
\frac{1}{2} \sum_{l=1}^{\infty} \langle \tilde{\Delta}_{n}
\tilde{\Delta}_{n-l} \rangle \cos \left( 2 k a l \right) .
\end{array}
\label{dethcorr}
\end{equation}

Plugging the noise-angle correlators~(\ref{uthcorr}) and~(\ref{dethcorr})
in Eq.~(\ref{lyap2}), one obtains that the Lyapunov exponent is
\begin{equation}
\lambda = \frac{1}{8a} \left[ \langle \tilde{u}_{n}^{2} \rangle
W_{1}(ka) + \langle \tilde{\Delta}_{n}^{2} \rangle W_{2}(ka) -
2 \langle \tilde{u}_{n} \tilde{\Delta}_{n} \rangle \cos \left( ka \right)
W_{3}(ka) \right]
\label{invloc}
\end{equation}
where the functions $W_{i}(ka)$, defined by the identities
\begin{equation}
\begin{array}{ccl}
W_{1} \left( ka \right) & = & \displaystyle
1 + 2 \sum_{l=1}^{\infty} \frac{\langle u_{n}u_{n+l} \rangle}
{ \langle u_{n}^{2} \rangle} \cos(2kal) \\
W_{2} \left( ka \right) & = & \displaystyle
1 + 2 \sum_{l=1}^{\infty} \frac{\langle \Delta_{n}\Delta_{n+l} \rangle}
{\langle \Delta_{n}^{2} \rangle} \cos (2kal) \\
W_{3} \left( ka \right) & = & \displaystyle
1 + 2 \sum_{l=1}^{\infty} \frac{\langle u_{n}\Delta_{n+l} \rangle}
{\langle u_{n} \Delta_{n} \rangle} \cos (2kal), \\
\end{array}
\label{powerspectra}
\end{equation}
are the Fourier transforms of the normalised binary
correlators~(\ref{bincor}).
Eq.~(\ref{invloc}), first derived in~\cite{Her08}, shows that the
Lyapunov exponent is the sum of three terms, with the first two
addends describing the effects of purely compositional and structural
disorder, while the third term is due to the interplay between these
two kinds of randomness.

We stress that expression~(\ref{invloc}) is valid for every value of
the energy within the allowed energy bands, with the exception of small
neighbourhoods at the band edges (i.e., for $ka \simeq 0$ and $ka \simeq
\pm \pi$) and close to the band centre (i.e., for $k \simeq \pi/2a$).

\section{Designed mobility edges}
\label{dme}

\subsection{Generation of self- and cross-correlated disorders}

In this section we discuss how specific long-range self-correlations of the
disorder can weaken or enhance the localisation of the electronic states
and produce pre-defined mobility edges. We then analyse the additional
effects of the cross-correlations between compositional and structural
disorders.
Formula~(\ref{invloc}) shows that the localisation length diverges in the
energy intervals where the power spectra~(\ref{powerspectra}) vanish.
The problem, therefore, is to generate two successions of random variables
$\{u_{n}\}$ and $\{\Delta_{n}\}$ with self- and cross-correlators such that
the power spectra~(\ref{powerspectra}) vanish in pre-assigned energy windows.
We are dealing with an ``inverse'' problem which, as such, has no
unique solution. Here we propose one, defining an algorithm for the
construction of two successions $\{u_{n}\}$ and $\{\Delta_{n}\}$ with the
required features.

As a first step, we consider two successions $\{X_{n}^{(1)}\}$ and
$\{X_{n}^{(2)}\}$ of independent random variables with zero mean and
unit variance. In other words, we require that
\begin{eqnarray*}
\langle X_{n}^{(i)} \rangle = 0 & \mbox{ and } &
\langle X_{n}^{(i)} X_{m}^{(j)} \rangle = \delta_{ij} \delta_{nm}
\end{eqnarray*}
for $i,j=1,2$ and $n,m \in {\bf Z}$.
In terms of these variables, we can construct two inter-correlated
successions
\begin{equation}
\begin{array}{ccl}
Y_{n}^{(1)} & = & X_{n}^{(1)} \cos \eta + X_{n}^{(2)} \sin \eta \\
Y_{n}^{(2)} & = & X_{n}^{(1)} \sin \eta + X_{n}^{(2)} \cos \eta \\
\end{array}
\label{iwn}
\end{equation}
with $\eta$ being a real parameter which determines the degree of
inter-correlation of the $Y$ variables.
In fact, one has
\begin{eqnarray*}
\langle Y_{n}^{(i)} \rangle = 0 & \mbox{ and } &
\langle Y_{n}^{(i)} Y_{m}^{(j)} \rangle = \delta_{nm} \left[
\delta_{ij} + \left( 1 - \delta_{ij} \right) \sin \left( 2 \eta \right)
\right]
\end{eqnarray*}
and this shows that the natural range of variation of the parameter $\eta$
is the interval $[-\pi/4, \pi/4]$ with $\eta = \pi/4$ corresponding to total
correlation of the $Y_{n}^{(1)}$ and $Y_{n}^{(2)}$ variables and
$\eta = -\pi/4$ to total anticorrelation (for $\eta = 0$ no cross-correlations
exist).

The next step consists in ``filtering'' the cross-correlated white-noise
successions $\{Y_{n}^{(1)}\}$ and $\{Y_{n}^{(2)}\}$ in order to obtain
two cross- and self-correlated sequences $\{u_{n}\}$ and $\{\Delta_{n}\}$.
For this purpose we express the $u_{n}$ and $\Delta_{n}$ random variables
as convolution products of the form
\begin{equation}
\begin{array}{ccc}
\displaystyle
u_{n} = \sum_{k=-\infty}^{\infty} \alpha_{k} Y_{n-k}^{(1)} &
\mbox{ and } &
\displaystyle
\Delta_{n} = \sum_{k=-\infty}^{\infty} \beta_{k} Y_{n-k}^{(2)} . \\
\end{array}
\label{convol}
\end{equation}
and we look for appropriate successions of the coefficients
$\alpha_{k}$ and $\beta_{k}$. We suppose that $\alpha_{k} = \alpha_{-k}$
and $\beta_{k} = \beta_{-k}$.
Starting from Eq.~(\ref{convol}) one obtains that the average values of the
variables $u_{n}$ and $\Delta_{n}$ vanish
\begin{eqnarray*}
\langle u_{n} \rangle = 0 & \mbox{ and } & \langle \Delta_{n} \rangle = 0
\end{eqnarray*}
and that the binary correlators take the values
\begin{eqnarray}
\langle u_{n} u_{n+k} \rangle & = &
\sum_{l=-\infty}^{\infty} \alpha_{l} \alpha_{l+k},
\label{ucorr} \\
\langle \Delta_{n} \Delta_{n+k} \rangle & = &
\sum_{l=-\infty}^{\infty} \beta_{l} \beta_{l+k},
\label{deltacorr} \\
\langle \Delta_{n} u_{n+k} \rangle & = & \langle u_{n} \Delta_{n+k} \rangle =
\sum_{l=-\infty}^{\infty} \alpha_{l} \beta_{l+k} \sin \left( 2 \eta \right).
\label{mixcorr}
\end{eqnarray}
Comparing Eqs.~(\ref{ucorr}) and~(\ref{deltacorr}) with the binary
self-correlators in Eq.~(\ref{bincor}), one arrives at the equations
\begin{eqnarray*}
\sum_{l=-\infty}^{\infty} \alpha_{l} \alpha_{l+k} & = &
\langle u_{n}^{2} \rangle \chi_{1}(k) \\
\sum_{l=-\infty}^{\infty} \beta_{l} \beta_{l+k} & = &
\langle \Delta_{n}^{2} \rangle \chi_{2}(k) \\
\end{eqnarray*}
whose solution is
\begin{equation}
\begin{array}{ccl}
\alpha_{k} & = & \displaystyle
\frac{2}{\pi} \int_{0}^{\pi/2} \sqrt{\langle u_{n}^{2} \rangle W_{1}(x)}
\cos \left( 2 k x \right) dx ,\\
\beta_{k} & = & \displaystyle
\frac{2}{\pi} \int_{0}^{\pi/2} \sqrt{\langle \Delta_{n}^{2} \rangle W_{2}(x)}
\cos \left( 2 k x \right) dx .\\
\end{array}
\label{ab}
\end{equation}
Inserting the coefficients~(\ref{ab}) in the convolution
products~(\ref{convol}) one obtains random variables $\{ u_{n} \}$
and $\{ \Delta_{n} \}$ with zero average and pre-assigned self-correlators
$\chi_{1}(k)$ and $\chi_{2}(k)$, as required.
To complete the picture, one should specify the inter-correlation parameter
$\eta$ which, together with the coefficients~(\ref{ab}) defines the
cross-correlator~(\ref{mixcorr}).

Thus, we have obtained a recipe that gives two random
successions $\{ u_{n} \}$ and $\{ \Delta_{n} \}$ in terms of the
pre-defined normalised self-correlators $\chi_{1}(k)$ and $\chi_{2}(k)$,
or, equivalently, of the pre-assigned power spectra $W_{1}(ka)$ and
$W_{2}(ka)$ and of the inter-correlation parameter $\eta$.
We can now see what is the form of the inverse localisation
length~(\ref{invloc}) when the random variables $u_{n}$ and $\Delta_{n}$
are generated with the above-described method.
After some algebra, one can write the cross power spectrum as
\begin{displaymath}
W_{3} (ka) = \frac{\displaystyle
\sqrt{\langle u_{n}^{2} \rangle W_{1}(ka) \langle \Delta_{n}^{2}
\rangle W_{2}(ka)}}
{\displaystyle \langle u_{n} \Delta_{n} \rangle} \sin \left( 2\eta \right).
\end{displaymath}
Inserting this result in expression~(\ref{invloc}) one obtains
\begin{equation}
\begin{array}{ccl}
\lambda & = & \displaystyle
\frac{1}{8a} \left[ \langle \tilde{u}_{n}^{2} \rangle
W_{1}(ka) + \langle \tilde{\Delta}_{n}^{2} \rangle W_{2}(ka) \right. \\
& - & \displaystyle \left.
2 \sqrt{\langle \tilde{u}_{n}^{2} \rangle
\langle \tilde{\Delta}_{n}^{2} \rangle W_{1}(ka) W_{2}(ka)} \;
\cos \left(ka \right) \sin \left( 2 \eta \right) \right] . \\
\end{array}
\label{ll}
\end{equation}

\subsection{Numerical examples}
\label{numex}

To test the validity of formula~(\ref{ll}) we construct a pair of random
sequences $\{ u_{n} \}$ and $\{ \Delta_{n} \}$ which enhance the
localisation of the states in two energy windows $[q_{1}^{2},q_{2}^{2}]$ and
$[q_{3}^{2},q_{4}^{2}]$ and delocalise the states in the rest of the band.
For the sake of simplicity we consider compositional and
structural disorders with identical self-correlators of the form
\begin{equation}
\chi_{1}(n) = \chi_{2}(n) = \frac{1}{2 \left(k_{2} - k_{1} \right)an}
\left[ \sin \left( 2 k_{2}a n \right) -
\sin \left( 2 k_{1}a n \right) \right]
\label{lrbincor}
\end{equation}
The power-law decay of this function is the signature of long-range
auto-correlations of the disorder.
The correlators~(\ref{lrbincor}) correspond to the power spectra
\begin{equation}
W_{1}(ka) = W_{2}(ka) = \left\{ \begin{array}{ccl}
\displaystyle
\frac{\pi}{2a \left(k_{2} - k_{1}\right)} &
\mbox{ if } & k \in \left[ k_{1}, k_{2} \right] \\
0 & \mbox{ if } &
k \in \left[ 0, k_{1} \right] \cup \left[ k_{2}, \frac{\pi}{2a}\right] \\
\end{array} . \right.
\label{powspe}
\end{equation}
In Eqs.~(\ref{lrbincor}) and~(\ref{powspe}), $k_{1}$ and $k_{2}$ represent
the two Bloch vectors corresponding to the mobility edges $q_{1}^{2}$ and
$q_{2}^{2}$, i.e.,
\begin{displaymath}
\begin{array}{ccc}
q(k_{1}) = q_{1} & \mbox{ and } & q(k_{2}) = q_{2} .
\end{array}
\end{displaymath}
Note that the power spectra~(\ref{powerspectra}) are periodic functions
of period $\pi/2a$ and therefore it is enough to define them in the
interval $[0, \pi/2a]$.

\begin{figure}[htp]
\begin{center}
\epsfig{file=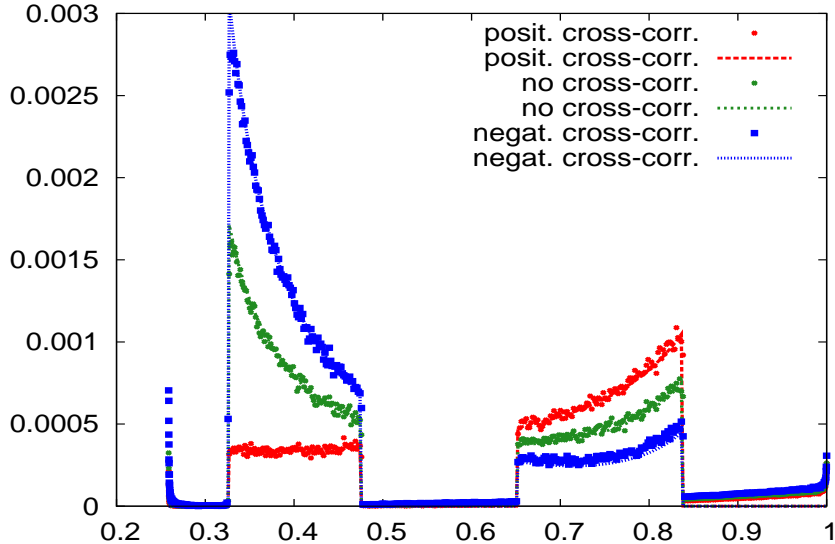,width=5in,height=3in}
\caption{Inverse localisation length versus $q/\pi$. Lines correspond to
the predictions of Eq.~(\ref{ll}); the symbols to numerical results.
\label{deloc1}}
\end{center}
\end{figure}

For our first numerical example we chose disorder strengths equal to
$\sqrt{\langle u_{n}^{2} \rangle} = \sqrt{\langle \Delta_{n}^{2} \rangle}
= 0.04$ and a mean field value $U = 0.7$.
We considered two Bloch vectors $k_{1} = \pi/5a$ and $k_{2} = 2\pi/5a$,
which, for the given value of $U$, correspond to mobility edges at
\begin{displaymath}
\begin{array}{cccc}
\tilde{q}_{1} = 0.327, & \tilde{q}_{2} = 0.476, & \tilde{q}_{3} = 0.652, &
\tilde{q}_{4} = 0.838 \\
\end{array}
\end{displaymath}
with $\tilde{q}_{i} = q_{i}/\pi$.
Note that, with our choice of $U$, the propagation wavevector $q$ within
the first energy bands spans the interval ranging from $q_{l}/\pi =  0.259$
to $q_{r}/\pi = 1.0$.
Fig.~\ref{deloc1} shows that the theoretical predictions of Eq.~(\ref{ll})
are well-matched by the numerical results.
As expected, the long-range self-correlations~(\ref{lrbincor}) produce sharp
mobility edges, whereas the cross-correlations can substantially change
the spatial extension of the localised states. In Fig.~(\ref{deloc1})
the case without cross-correlations ($\eta = 0$) is compared with the two
extreme cases of total positive ($\eta = \pi/4$) and negative
($\eta = -\pi/4$) cross-correlations.

\begin{figure}[htp]
\begin{center}
\epsfig{file=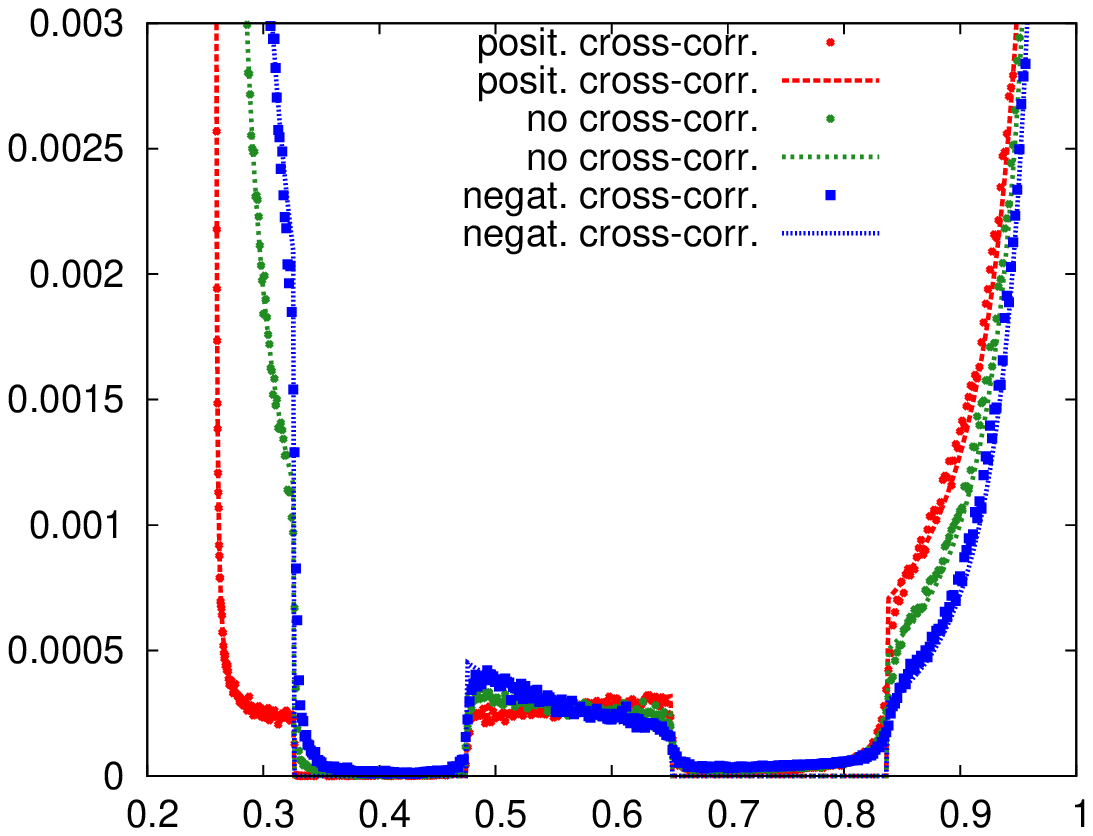,width=5in,height=3in}
\caption{Inverse localisation length versus $q/\pi$. Lines correspond to
theoretical predictions and symbols to numerical results.\label{deloc2}}
\end{center}
\end{figure}

As a second example, in Fig.~\ref{deloc2} we represent the theoretical
predictions and numerical results for the complementary case,
characterised by the same mobility edges of the previous example but 
with inverted windows of localised and delocalised states.
(The values of the disorder strengths and of the mean field are the
same as in the previous case.)
We observe again a good correspondence between the numerical values and the
inverse localisation length~(\ref{ll}).

\begin{figure}[htp]
\begin{center}
\epsfig{file=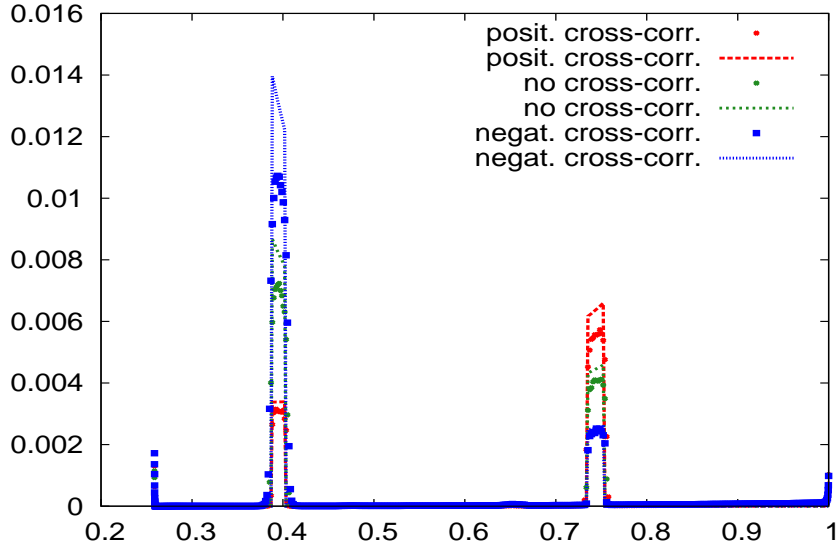,width=5in,height=3in}
\caption{Inverse localisation length versus $q/\pi$. Lines correspond to
theoretical predictions and symbols to numerical results. \label{enhanloc}}
\end{center}
\end{figure}

To conclude this section, we observe that the spatial extension of the
localised states can be significantly reduced by squeezing the energy
windows of localised states.
This localisation enhancement, first discussed in~\cite{Kuh08}, is a
consequence of the normalisation condition
\begin{displaymath}
\int_{0}^{\pi/(2a)} W_{i}(ka) dk = \frac{\pi}{2a},
\end{displaymath}
which follows from the fact that $\chi_{i}(0) = 1$.
To illustrate this effect, we consider a third example, with
all parameters unchanged with respect to the first one ($U = 0.7$,
$\sqrt{\langle u_{n}^{2} \rangle} = \sqrt{\langle \Delta_{n}^{2} \rangle}
= 0.04$) except for the size of the windows of localised states, which we
shrink by setting $k_{1} = 0.29 \pi/a$ and $k_{2} = 0.31 \pi/a$.
In terms of the propagation wavevector, these values of the Bloch vectors
correspond to mobility edges at
\begin{equation}
\begin{array}{cccc}
\tilde{q}_{1} = 0.388, & \tilde{q}_{2} = 0.402, & \tilde{q}_{3} = 0.736, &
\tilde{q}_{4} = 0.755 .\\
\end{array}
\label{mobedges}
\end{equation}
The theoretical and numerical results for the inverse localisation length,
represented in Fig.~\ref{enhanloc}, display a clear enhancement of
localisation in the narrowed energy intervals of localised
states. The stronger localisation magnifies the effect of the
cross-correlations of the compositional and structural disorder,
enlarging the difference between the extreme cases of total positive
and negative inter-correlations.

\section{Transport properties}
\label{trans}

In the previous sections we have considered the structure of the electronic
states in a infinite Kronig-Penney model. We now turn our attention
to the relevant problem of electronic transmission through finite
disordered segments.
We consider the case of a random Kronig-Penney model extending over $N$
lattice sites sandwiched between two semi-infinite perfect leads.
From the mathematical point of view, this means that the variables $u_{n}$
and $\Delta_{n}$ in the Schr\"{o}dinger equation~(\ref{kpmodel}) are defined
as before for $1 \leq n \leq N$, but vanish for $n < 1$ and $n > N$. 
We assume that the left lead carries an incoming and a reflected wave
with wavevector $q$, 
\begin{displaymath}
\begin{array}{lcc}
\psi_{n} = e^{iqan} + r e^{-iqan} & \mbox{ for } & n = 0, -1, -2, \ldots
\end{array}
\end{displaymath}
while a transmitted wave propagates in the right lead
\begin{displaymath}
\begin{array}{cll}
\psi_{n} = t e^{iqan} & \mbox{ for } & n = N+1, N+2, \ldots
\end{array}
\end{displaymath}

To determine the transmission coefficient $T(q) = |t(q)|^{2}$ we
follow the variant of the transfer-matrix approach introduced in~\cite{Kot97}.
After eliminating the momenta from the map~(\ref{hammap1}), with the
obvious substitution $x_{n} \rightarrow \psi_{n}$ we obtain the equation
\begin{equation}
\begin{array}{cl}
& \displaystyle
\frac{1}{\sin\left[ q \left(a + \Delta_{n} \right) \right]} \psi_{n+1} +
\frac{1}{\sin\left[ q \left(a + \Delta_{n-1} \right) \right]} \psi_{n-1} \\
= & \displaystyle
\left\{ \cot \left[ q \left(a + \Delta_{n} \right) \right] +
\cot \left[ q \left(a + \Delta_{n-1} \right) \right] + \frac{1}{q}
\left( U + u_{n} \right) \right\} \psi_{n} ,
\end{array}
\label{tbmodel}
\end{equation}
which defines the tight-binding system corresponding to the Kronig-Penney
model~(\ref{kpmodel}). Note that the compositional disorder in the
Kronig-Penney model manifests itself as diagonal disorder in
Eq.~(\ref{tbmodel}), whereas the structural disorder shows up in both the
diagonal and off-diagonal random coefficients of the tight-binding
model~(\ref{tbmodel}).
Using the transfer-matrix formalism, one can write Eq.~(\ref{tbmodel}) in
the form
\begin{equation}
\left( \begin{array}{c}
\psi_{n+1} \\
\psi_{n} \\
\end{array} \right)
= {\bf M}_{n}
\left( \begin{array}{c}
\psi_{n} \\
\psi_{n-1} \\
\end{array} \right)
\label{tmmap}
\end{equation}
with the elements of the matrix ${\bf M}_{n}$ being
\begin{displaymath}
\begin{array}{ccl}
\left( {\bf M}_{n} \right)_{11} & = &
\left\{ \cot \left[ q \left( a + \Delta_{n} \right) \right] +
\cot \left[ q \left( a + \Delta_{n-1} \right) \right] +
\frac{1}{q} \left( U + u_{n} \right) \right\}
\sin \left[ q \left( a + \Delta_{n} \right) \right], \\
\left( {\bf M}_{n} \right)_{12} & = & \displaystyle
-\frac{\sin \left[ q \left( a + \Delta_{n} \right) \right]}
{\sin \left[ q \left( a + \Delta_{n-1} \right) \right]}, \\
\left( {\bf M}_{n} \right)_{21} & = & 1, \\
\left( {\bf M}_{n} \right)_{22} & = & 0. \\
\end{array}
\end{displaymath}
The total transfer matrix connecting the wavefunctions in the two
leads is then
\begin{equation}
{\bf M} = {\bf M}_{N} {\bf M}_{N-1} \cdots {\bf M}_{1} .
\label{totaltm}
\end{equation}
Following Pichard~\cite{Pic86}, one can write the transmission
coefficient in terms of the elements of the matrix~(\ref{totaltm}).
One thus obtains the expression
\begin{equation}
T_{N}(q) = \frac{4 \sin^{2}(q) \left( \det {\bf M} \right)^{2}}
{\left| {\bf M}_{21} - {\bf M}_{12} + {\bf M}_{22} e^{iq} -
{\bf M}_{11} e^{-iq} \right|^{2}} .
\label{pichard}
\end{equation}
We remark that the factor $(\det {\bf M})^{2}$ enters the previous
formula because the determinant of the transfer matrices ${\bf M}_{n}$
is not unitary. More precisely, one has
\begin{displaymath}
\det {\bf M} = \frac{\sin \left[ q \left( a + \Delta_{N} \right) \right]}
{\sin \left[ q \left( a + \Delta_{0} \right) \right]} .
\end{displaymath}
To avoid the nuisance of a total transfer matrix with non-unitary
determinant, one can redefine the random barrier as including an
extra site with no disorder on each side. From now on we will adopt this
convention and drop the determinant factor in formulae derived from
Eq.~(\ref{pichard}).

To express the transmission coefficient~(\ref{pichard}) in a more convenient
way, we perform a similarity transformation
\begin{displaymath}
{\bf Q}_{n} = {\bf R} {\bf M}_{n} {\bf R}^{-1}
\end{displaymath}
with
\begin{displaymath}
{\bf R} = \left( \begin{array}{cc}
1 & 0 \\
\displaystyle
\cos \left( qa \right)/\sin \left( qa \right) &
\displaystyle
-1/\sin \left( qa \right) \\
\end{array} \right)
\end{displaymath}
and put the map~(\ref{tmmap}) in the equivalent form
\begin{equation}
\left( \begin{array}{c}
x_{n+1} \\
p_{n+1} \\
\end{array} \right) =
{\bf Q}_{n}
\left( \begin{array}{c}
x_{n} \\
p_{n} \\
\end{array} \right) .
\label{qmap}
\end{equation}
This allows one to cast the transmission coefficient~(\ref{pichard})
in the more appealing form~\cite{Kot97}
\begin{equation}
T_{N}(q) = \frac{4}{2 + {\bf v}_{1} \cdot {\bf v}_{1} +
{\bf v}_{2} \cdot {\bf v}_{2}} ,
\label{trampl}
\end{equation}
with ${\bf v}_{1}$ and ${\bf v}_{2}$ being the two-component vectors
defined as
\begin{displaymath}
{\bf v}_{1} = {\bf Q}_{N+1} {\bf Q}_{N} \cdots {\bf Q}_{1} {\bf Q}_{0}
\left( \begin{array}{c}
1 \\ 0 \\
\end{array} \right)
\end{displaymath}
and
\begin{displaymath}
{\bf v}_{2} = {\bf Q}_{N+1} {\bf Q}_{N} \cdots {\bf Q}_{1} {\bf Q}_{0}
\left( \begin{array}{c}
0 \\ 1 \\
\end{array} \right) .
\end{displaymath}
Eq.~(\ref{trampl}) is very convenient for the numerical evaluation of
the transmission coefficient, which can be efficiently determined by
computing the evolution of the initial vectors $(1,0)$ and $(0,1)$
under the map~(\ref{qmap}).

Using this approach, we numerically investigated the effects of the
long-range correlations of the disorder on the transmission
coefficient $T_{N}(q)$.
In the absence of cross-correlations, only two length scales are relevant
for the transport properties of the disordered segment, namely, the length
of the segment itself, $L = Na$ and the spatial extension $l_{\rm loc}$ of
the localised states in the infinite random model.
In principle one should also consider the localisation length
$l_{\rm del}$ of the delocalised states, because the vanishing of the
second-order Lyapunov exponent~(\ref{invloc}) does not necessarily
imply that the higher-order terms must also be zero.
However, for weak disorder the spatial extension of the delocalised
states is so large that it can be considered as infinite with respect
to the length $L$ of the random barrier.

In the localised regime, i.e., when the condition $l_{\rm loc} \ll L$
is fulfilled, our numerical experiments show that the windows of
delocalised states created by long-ranged self-correlations of the
disorder survive in the case of disordered segments, where they manifest
themselves as energy windows with transmittivity close to one.
This agrees with the results already obtained, both numerically and
experimentally, for barriers with purely compositional or structural
disorder~\cite{Izr01,Lun09}.
Our numerical data also show that the cross-correlations of the
disorder have additional effects on the transport properties.
Their influence is most evident in the localisation windows created by
long-ranged self-correlations of the disorder, where cross-correlations
either enhance or diminish the electronic transmission according to whether
they increase or decrease the localisation length.

\begin{figure}[htp]
\begin{center}
\epsfig{file=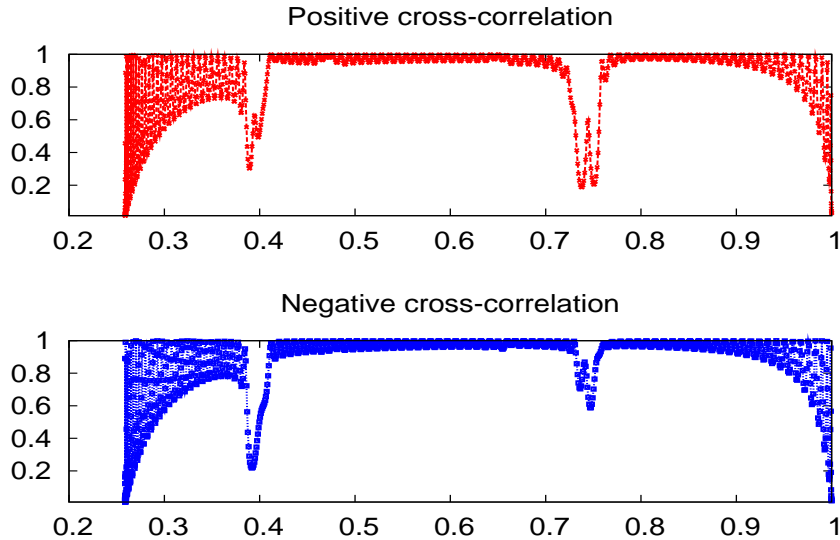,width=5in,height=3in}
\caption{Transmission coefficient $T_{N}(q)$ versus $q/\pi$ for
random sample of $N = 100$ sites.
\label{ta_n100}}
\end{center}
\end{figure}

\begin{figure}[htp]
\begin{center}
\epsfig{file=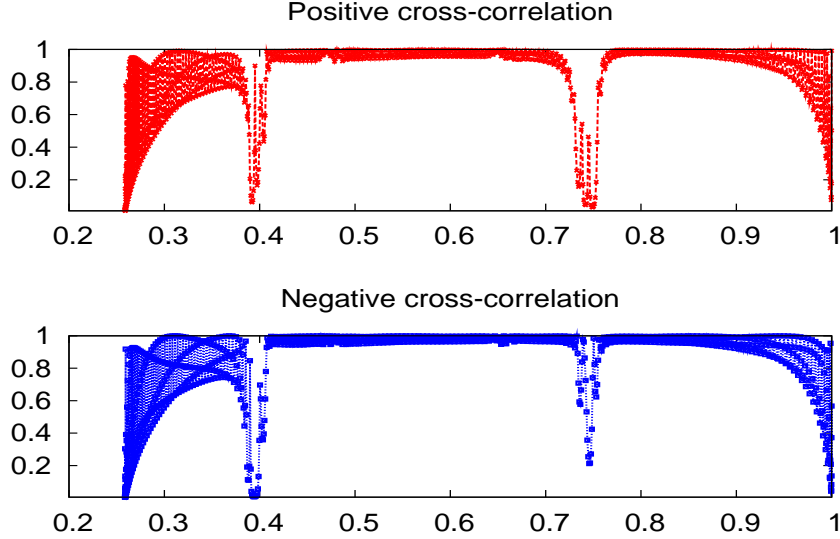,width=5in,height=3in}
\caption{Transmission coefficient $T_{N}(q)$ versus $q/\pi$ for
random sample of $N = 200$ sites.
\label{ta_n200}}
\end{center}
\end{figure}

The effects of both the self- and the cross-correlations of the disorder
can be appreciated in Figs.~\ref{ta_n100} and~\ref{ta_n200}, which
represent the transmission amplitude for two lengths of a random sample
with the same disorder characteristics of the third example analysed in
Sec.~\ref{numex}.
In other words, we consider a random Kronig-Penney barrier with
mean field $U = 0.7$, disorder strengths $\sqrt{\langle u_{n}^{2} \rangle}
= \sqrt{\langle \Delta_{n}^{2} \rangle} = 0.04$, and windows of localised
states $[q_{1},q_{2}]$ and $[q_{3},q_{4}]$, with edges defined by
Eq.~(\ref{mobedges}).
Each figure represents the transmission coefficients for the cases
of total positive and negative cross-correlations.

When the cross-correlations are taken into account, it is better to
characterise the spatial extension of the localised states in terms not
of one but of two length scales, i.e., $l_{\rm loc}^{+}(q)$ and
$l_{\rm loc}^{-}(q)$, which correspond to the maximum and minimum values
of the localisation length obtained by varying the inter-correlation of
the compositional and structural disorders.
In the present case, one has $l_{\rm loc}^{-} \sim 100 a$ and
$l_{\rm loc}^{+} \sim 300 a$ in the localisation window of lower energy
and $l_{\rm loc}^{-} \sim 200 a$ and $l_{\rm loc}^{+} \sim 400 a$ in the 
higher-energy localisation window.
The data represented in Figs.~\ref{ta_n100} and~\ref{ta_n200} were obtained
for the sample lengths $L = 100a$ and $L = 200a$, which correspond to
the intermediate region between the ballistic regime, characterised by
$L \ll l_{\rm loc}^{-}$ and the exponentially localised regime for which
$l_{\rm loc}^{+} \ll L$.

Two main features emerge from the numerical data. On the one hand, as $L$
is increased the mobility edges become sharper and sharper, because the
localisation of the electronic state in the windows $[q_{1},q_{2}]$ and
$[q_{3},q_{4}]$ becomes more and more effective.
On the other hand, the existence of two localisation lengths implies that
the regime of exponential localisation is attained with different speed
according to the type and degree of cross-correlations.
This is the physical origin of the difference in the transmission
coefficients for positive and negative cross-correlations.

In general terms, one should notice that, when the length of the sample
is less than the minimum localisation length $l_{\rm loc}^{-}$, the
localisation of the electronic states is very weak, and this can produce
strong sample-to-sample fluctuations of the transmission coefficient
which partially mask the effects of cross-correlations.
As the length of the barrier is increased, these effects become clearly
discernible and their influence peaks in the intermediate region
between the ballistic and the localised regimes.
If one lets the length of the random sample grow further, the regime of
exponential localisation is eventually reached for both kinds of
cross-correlations. In this regime both transmission coefficients are
exponentially small in the localisation windows and their absolute
difference is therefore hard to detect.

\section{Conclusions}
\label{conclu}

In this paper we have analysed the structure of the electronic states and
transport properties of a Kronig-Penney model with weak compositional
and structural disorder. We have shown that specific long-range
self-correlations of the disorder can enhance or suppress the localisation
of the electronic states in pre-defined energy windows and that an
additional modulation of the localisation length can be obtained by
cross-correlating the two kinds of disorder. A method to generate
compositional and structural disorder with arbitrary self- and
cross-correlations has been given.

The analysis performed in the second part of this work shows that, by
carefully selecting the statistical properties of the disorder, one can
produce random samples with peculiar transport properties. In particular,
it is possible to obtain almost perfect transmission in pre-defined energy
windows. Although self-correlations play a dominant role in shaping the
transport properties of finite-size random chains, cross-correlations
can be used to fine-tune the transmission coefficient.

J. C. H.-H. and L.T. gratefully acknowledge the support of the CONACyT grant
nr. 84604 and of the CIC-2009 grant (Universidad Michoacana).
The work of F.M.I. was partly supported by the CONACyT grant nr. 80715.

\end{document}